\documentstyle[aps,twocolumn,epsfig]{revtex}

\begin{document}

\draft
\title{Frequency Shift and Mode Coupling in the \\
Nonlinear Dynamics of a Bose Condensed Gas }
\author{F. Dalfovo$^1$, C. Minniti$^1$ and L.P. Pitaevskii$^{1,2,3}$}
\address{$^1$ Dipartimento di Fisica, Universit\`a di Trento, and \\
Istituto Nazionale Fisica della Materia, I-38050 Povo, Italy}

\address{$^2$ Department of Physics, TECHNION, Haifa 32000, Israel}

\address{$^3$ Kapitza Institute for Physical Problems, Kosygina 2, 
117334 Moscow}

\date{August 13, 1997}

\maketitle

\begin{abstract}
We investigate the behavior of  large amplitude oscillations of
a trapped Bose-condensed gas of alkali atoms at zero temperature, 
by solving the equations of hydrodynamics for collective modes. 
Due to the atom-atom interaction, the equations of motion
are nonlinear and give rise to significant frequency shift and mode 
coupling.   We provide analytic expressions for the frequency 
shift, pointing out the  crucial role played by the anisotropy of the 
confining potential. For special values of the anisotropy parameter the 
mode coupling is particularly strong and the frequency shift becomes 
large, revealing a peculiar behavior of the Bose-condensed gas. 
Consequences on the theory of collapse and revival of collective
excitations are also discussed. 
\end{abstract}

\pacs{3.75.Fi, 67.40.Db}

\narrowtext

\section{Introduction}
\label{sec:introduction}

One of the most important features of an interacting 
quantum many-body system is its response to external oscillating 
fields.  The collective modes, which are expected to dominate the
low frequency response, represent a 
very effective tool  for probing the role of interactions 
and testing theoretical schemes. For this reason, measurements 
of collective modes in the trapped gases of alkali atoms \cite{Jila,MIT} 
were carried out soon after the discovery of 
Bose-Einstein condensation \cite{discovery}. The remarkable 
agreement between measured frequencies and theoretical predictions 
\cite{Edwards,Stringari,Singh,Victor,Esry,You} is one of the first 
important achievements in the investigation of these new 
systems. It provides also a clean {\it a posteriori}
justification of the mean-field scheme, based on the 
formalism  of Gross-Pitaevskii theory \cite{GP},  which is the 
starting point of most calculations. This theoretical approach 
is  expected to give indeed an accurate description of the ground 
state and the  excited states of such dilute interacting gases 
at low temperature. The same scheme, when the atom-atom interaction 
becomes dominant  compared with the zero-point quantum pressure, 
reduces to the Thomas-Fermi approximation for the ground state 
and to the equations of nondissipative hydrodynamics for the excited 
states \cite{Stringari,Fetter,Wu}. This is particularly useful for
discussing the relevant physical properties by means of analytic 
or semi-analytic results, taking advantage of the simplicity 
and clarity of the hydrodynamic equations. 

The same mean-field approach, which correctly reproduces the frequency
of the normal modes of the trapped gas in the linear limit (small
oscillations around the ground state), is suitable to investigating 
also the nonlinear dynamics of the systems
\cite{Ruprecht,Holland,Castin,Kagan,TN,Smerzi,Ohberg,Miroslaw}. 
The nonlinearity is  included in the equations of motion through the 
mean-field, which is proportional to the condensate density. Thus, 
measurable  effects of nonlinearity could represent further clean 
signatures of Bose-Einstein condensation. Among them, large amplitude 
oscillations of the condensate can be easily produced in the trapped gases
\cite{Jila,MIT}; nonlinear effects are expected to give rise to frequency
shift, mode coupling, harmonic generation and stochastic motion.  

The purpose of the present work is to derive simple differential
equations for large amplitude oscillations within the formalism of 
hydrodynamics. In particular we will provide  
analytic formulae for the frequency shift of three collective 
modes in a generic axially-symmetric trap. These formulae allow 
us to discuss the important role played by the anisotropy of the trap. 
We will show that special  values of the anisotropy parameter (the ratio 
of the  axial and radial frequencies of the trap) can be associated with
strong nonlinear effects even for oscillations of relatively small 
amplitude. The same analytic results have interesting consequences
in the theory of collapse and revival of the condensate. 

In the next section we introduce the basic formalism. Then, 
in section \ref{sec:oscillations}, we discuss how the collective
modes can be driven and analysed numerically. In section 
\ref{sec:analytic} we perform a small amplitudes expansion
and derive analytic solutions for the $m=0$ and $m=2$ modes
and their frequency shifts. In section \ref{sec:results} we
discuss both the numerical and the analytic results for different 
traps.  An application to the theory of collapse and revival of the
the oscillations is given in section \ref{sec:collapse}. The 
paper will end with a short summary. 

\section{Basic formalism}
\label{sec:basic}

Let us start with the hydrodynamic equations in Thomas-Fermi
approximation \cite{Stringari}: 
\begin{equation}
\frac{\partial}{\partial t} \rho + {\bf \nabla} 
\cdot ({\bf v}\rho) = 0 
\label{continuity} 
\end{equation}
\begin{equation}
m \frac{\partial}{\partial t} {\bf v} +
{\bf \nabla} \left( V_{ext} + g\rho 
+ { mv^2 \over 2 } \right) = 0 \; .
\label{Euler} 
\end{equation}
Density and velocity are related to the condensate wave function 
$\Psi({\bf r},t)$ through $\rho=|\Psi|^2$ and 
${\bf v}= \hbar(2mi\rho)^{-1} (\Psi^* {\bf \nabla}\Psi 
-\Psi{\bf \nabla}\Psi^*)$. The density is normalized to the number of 
particles in the condensate, $N=\int d{\bf r} \rho ({\bf r})$. 
The external confining potential has the
form $V_{ext}({\bf r})= (1/2) m \sum_i  \omega_{i}^2 r_i^2$, where 
$r_i \equiv x,y,z$. The trapping frequencies can depend on time,
$\omega_{i}=\omega_{i}(t)$,  in the presence of an external driving 
force. Their static values, $\omega_{0i}=\omega_{i}(0)$, fix the 
equilibrium configuration of the system.  
For cylindrically symmetric traps one can 
define the radial frequency $\omega_\perp \equiv \omega_{0x}=
\omega_{0y}$ and  the  asymmetry parameter $\lambda=\omega_{0z}/
\omega_\perp$. The external potential provides also the typical length 
scale of the system in each direction, $a_{HO}^{(i)}= 
\sqrt{\hbar /(m\omega_{0i})}$. Since 
the system is dilute, the atom-atom interaction enters only through 
the quantity $g=4\pi \hbar^2 a/m$, where $a$ is the  $s$-wave 
scattering length. 

The equations of nondissipative hydrodynamics are equivalent to the 
time dependent Gross-Pitaevskii equation for the condensate 
wave function in the limit $Na/a_{HO}\gg 1$, if the interaction
is repulsive ($a > 0$). In this
case, the effects of the zero point kinetic energy (quantum 
pressure) become negligible and
the gas is dominated by the balance of the internal and external 
potential energies. The stationary solution of the hydrodynamic
equations is the Thomas-Fermi ground state 
\cite{Stringari,Baym}:
\begin{equation}
\rho_0^{TF}({\bf r}) = | \Psi_0^{TF} ({\bf r})|^2 
= g^{-1} [ \mu - V_{ext}({\bf r}) ] \; \; \; \hbox{for} \; \; \; 
\mu \ge V_{ext}({\bf r})  
\label{tfgs}
\end{equation}
where the  chemical potential $\mu$ is fixed by the normalization 
of the density to the number of particle $N$.  The hydrodynamic
approach works in an excellent way for the lowest collective
modes of the Sodium atoms trapped at MIT \cite{MIT}, where $N$ is 
of the order of 1 million and more. Conversely, it provides only 
a semi-quantitative description of the Rubidium gas first trapped 
at Jila \cite{Jila}, where the number of atoms was smaller 
($10^3$-$10^4$).  Even in that case, however, the measured frequencies
converge nicely to the hydrodynamic predictions for the largest 
values of $N$. Compared with the numerical solution of the 
Gross-Pitaevskii equation, the hydrodynamic 
formalism has the advantage of providing analytic results for 
the dispersion law of the collective modes and for other useful
quantities. 

As already discussed in our previous paper \cite{TN}, exact 
solutions  of the hydrodynamic equations can be found in the 
form 
\begin{eqnarray}
\rho({\bf r},t) &=& a_x(t)  x^2 + a_y(t) y^2 + a_z(t) z^2 + a_0(t)
\label{scalingrho} \\
{\bf v}  &=& {1 \over 2} {\bf \nabla} [\alpha_x(t) x^2 + 
\alpha_y(t) y^2 + \alpha_z(t) z^2] \; ,
\label{scalingv}
\end{eqnarray}
restricted to the region where $\rho\ge 0$. With this choice, 
equations (\ref{continuity}-\ref{Euler}) transform into a set 
of coupled differential equations for the time-dependent 
coefficients $a_j(t)$ and $\alpha_j(t)$. One of them is fixed 
by the conservation of $N$:  $a_0=-(15N/8\pi)^{2/5}(a_xa_ya_z)^{1/5}$. 
The equations for the others can be further simplified by introducing 
the new variables $b_i$ defined by  
$a_i=-m\omega_{0i}^2 (2g b_x b_y b_z
b_i^2)^{-1}$. The hydrodynamic equations then yield 
$\alpha_i=\dot{b}_i/b_i$ and 
\begin{equation}
\ddot{b}_i  + \omega_{i}^2 b_i - \omega_{0i}^2 /(b_i b_x b_y b_z) 
= 0 \; ,
\label{ddotb}
\end{equation}
with $i=x,y,z$. These equations describe  the time evolution  
of the widths of the atomic cloud, since the new variables 
$b_i$ are directly related to the  mean square radii and velocities of 
the system \cite{TN}: $b_i^2 \propto \langle r^2_i \rangle$ and 
$\dot{b}_i^2 \propto \langle v^2_i \rangle$.  Different derivations 
of equations (\ref{ddotb}) and some  applications are given  in 
Refs.~\cite{Castin,Kagan,TN}. A variational approach including 
the  zero point quantum pressure, beyond the Thomas-Fermi 
approximation, has been also presented in Ref.~\cite{Victor}. 
Note again that the 
frequencies $\omega_i$, entering the second term of (\ref{ddotb}),  
can depend on time and, hence, these equations can be used for 
describing time varying traps, as well as the expansion of the
gas after a sudden switching-off of the confining potential. 
The set of solutions defined by the scaling transformations 
(\ref{scalingrho}-\ref{scalingv}) does not exhaust all possible 
motions of the trapped gas. For instance, the motion of the 
center of mass can be included by adding terms linear in  $x$, 
$y$ or  $z$ and  other solutions can be found including terms 
of the form $xy$, $xz$ or $yz$. However, they are well suited to 
study the collective modes of lowest multipolarity and energy,
namely the $m=0$ and $m=2$ modes, where $m$ is the azimuthal 
angular momentum in the cylindrically symmetric trap. In the
following we will apply equation (\ref{ddotb}) to these modes
in the nonlinear regime.

\section{Oscillations of a driven condensate}
\label{sec:oscillations}

At equilibrium one has  $b_i=1$ and $\dot{b}_i=0$. One can 
perturb the system by modulating the trap frequencies for 
a certain time and then let it oscillate freely. Formally,
this means that equations (\ref{ddotb}) have to be solved
using a time dependent frequency of the form  
$\omega_i^2(t)=\omega_{0i}^2 [1 + 2 f_i(t)]$, where $f_i(t)$
is an appropriate sinusoidal function within a finite time
interval, while the factor $2$ is introduced for 
convenience. An example is given in Fig.~1
for the case of a trap with asymmetry parameter 
$\lambda= \omega_{0z}/\omega_\perp = \sqrt{8}$, as in 
Jila experiments. Starting from equilibrium, an oscillation
is induced by choosing $f_x(t) = \eta \sin \Omega_d t$, 
$f_y = -f_x$ and $f_z=0$, with a driving frequency 
$\Omega_d = \sqrt{2} \ \omega_\perp$.  The driving force is 
switched-off  at $t=20 \omega_{\perp}^{-1}$. The three curves 
correspond to  the width $b_x(t)$ plotted for different values 
of the  driving strength, $\eta$. The free oscillations 
are undamped, since the theory is restricted to zero 
temperature and does not include any dissipation. 
In the small amplitude limit, $\eta \to 0$,
the oscillation coincides with the $m=2$ normal mode predicted
by the linearized hydrodynamic equations \cite{Stringari}.
It corresponds to a quadrupole-type excitation in the $xy$-plane. 
For larger amplitudes, the response of the system is slightly 
shifted in frequency, as can be seen from the figure,
and the oscillations are no more purely sinusoidal. This is 
associated with the occurrence of harmonic generation and 
mode coupling. 
In a similar way, one can excite the $m=0$ modes. The one at low 
energy corresponds to an in-phase oscillation of the width along 
$x$ and $y$ and out-of-phase along $z$. The one at high energy
is an in-phase compressional mode along all directions (breathing
mode). In all cases, one can solve numerically the equations 
(\ref{ddotb}) and extract amplitude and frequency of the 
excited modes. A simple way consists in doing a best-fit with 
a sinusoidal function to the appropriate width $b_i(t)$, after 
the switching-off of the driving force.  One can also perform a 
Fourier analysis of the signal. We used both methods, finding the
same results provided the time interval contains at least three or 
four complete oscillations. 
\begin{figure}
\begin{center}
  \epsfig{file=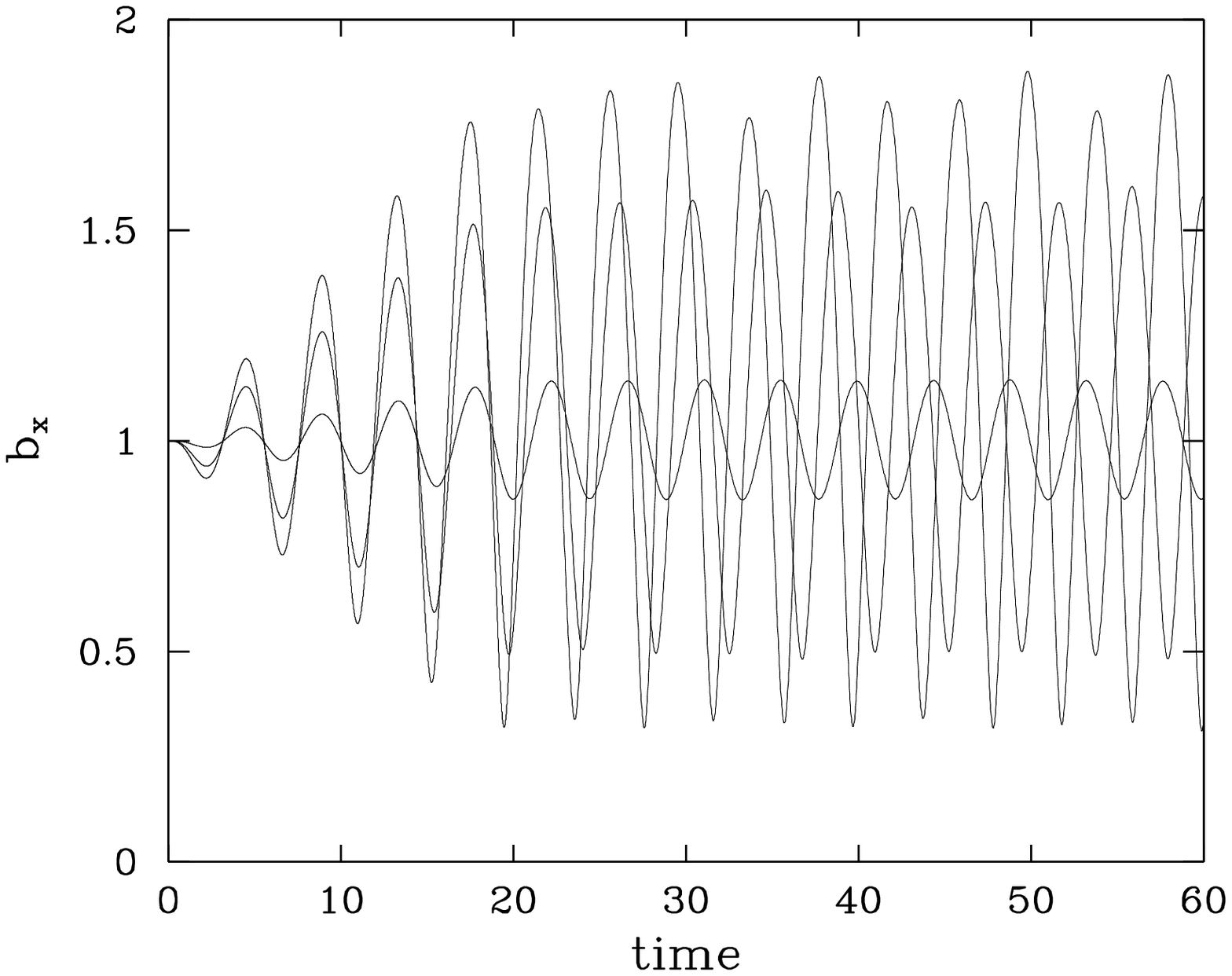,width=\linewidth}
  \begin{caption}
{  
 Evolution in time of the radial width $b_x(t)$ in a trap with
$\lambda=\protect\sqrt{8}$. Time is in units of $\omega_\perp^{-1}$. The 
oscillation  is driven by an external force for the first $20$ time 
units; then the condensate oscillates freely. The three curves 
correspond to increasing the strength of the driving force 
($\eta=0.01, 0.04, 0.06$, see text). In the small amplitude limit,
the oscillation coincides with the $m=2$ normal mode.  }
\end{caption}
\end{center}
\label{fig:m2-jila-eta}
\end{figure}

In order to point out the effects of mode-coupling, it is convenient to 
introduce suitable combinations of the widths $b_i(t)$ as follows:
\begin{eqnarray}
b_{x}(t) & = & 1 + \xi_-(t) + \xi_+(t) + \xi_{2}(t) 
\label{bx} \\ 
b_{y}(t) & = & 1 + \xi_-(t) + \xi_+(t) - \xi_{2}(t)   
\label{by} \\ 
b_{z}(t) & = & 1 + (q_{-}-4)\ \xi_-(t) + (q_{+}-4)\ \xi_+(t)  \; ,
\label{bz}
\end{eqnarray}
with 
\begin{equation}
q_{\mp} = 2+ (3/2)  \lambda^{2} \mp (1/2) 
\sqrt{ 9 \lambda^{4}-16 \lambda^{2}+16 } \; . 
\label{qmp}
\end{equation}
Inserting these combinations into (\ref{ddotb}) one obtains equations
for $\xi_-, \xi_+$ and $\xi_{2}$. Such equations are completely 
decoupled in the linear limit, that is, when 
the perturbations $\xi_-, \xi_+$ and $\xi_{2}$ are much smaller 
than $1$. Thus, these functions represent the normal modes of the system:
$\xi_{2}$ is the $m=2$ mode, having frequency $\sqrt{2}\ \omega_\perp$, 
while $\xi_-$ and $\xi_+$ are the  low-lying and high-lying 
$m=0$ modes, respectively, whose frequencies are $\sqrt{q_{\mp}}\ 
\omega_\perp$ \cite{Stringari}. We note also that $q_+ \ne q_-$ for
any value of $\lambda$. 

In Fig.~2 we plot the functions $\xi_j(t)$ for the
oscillation of largest amplitude already shown in  
Fig.~1. The driving force is tuned on  
the $m=2$ mode, $\xi_2$, but, due to nonlinear coupling, the
low-lying $m=0$ mode,  $\xi_-$, is also significantly excited,
while $\xi_+$ remains very small. The coupling between these modes 
causes the irregular oscillation of $b_x$, given by (\ref{bx}) and
plotted in  Fig.~1. These oscillations have
a very large amplitude (about $80$\% of the radial width); for
smaller oscillations the coupling with $\xi_-$ tends to vanish. 
For the smallest oscillation in Fig.~1 both
$\xi_-$ and $\xi_+$ are negligible with respect to the driven 
mode $\xi_2$ and one has also  $b_x(t) \simeq 1+\xi_2(t)$.
\begin{figure}
\begin{center}
  \epsfig{file=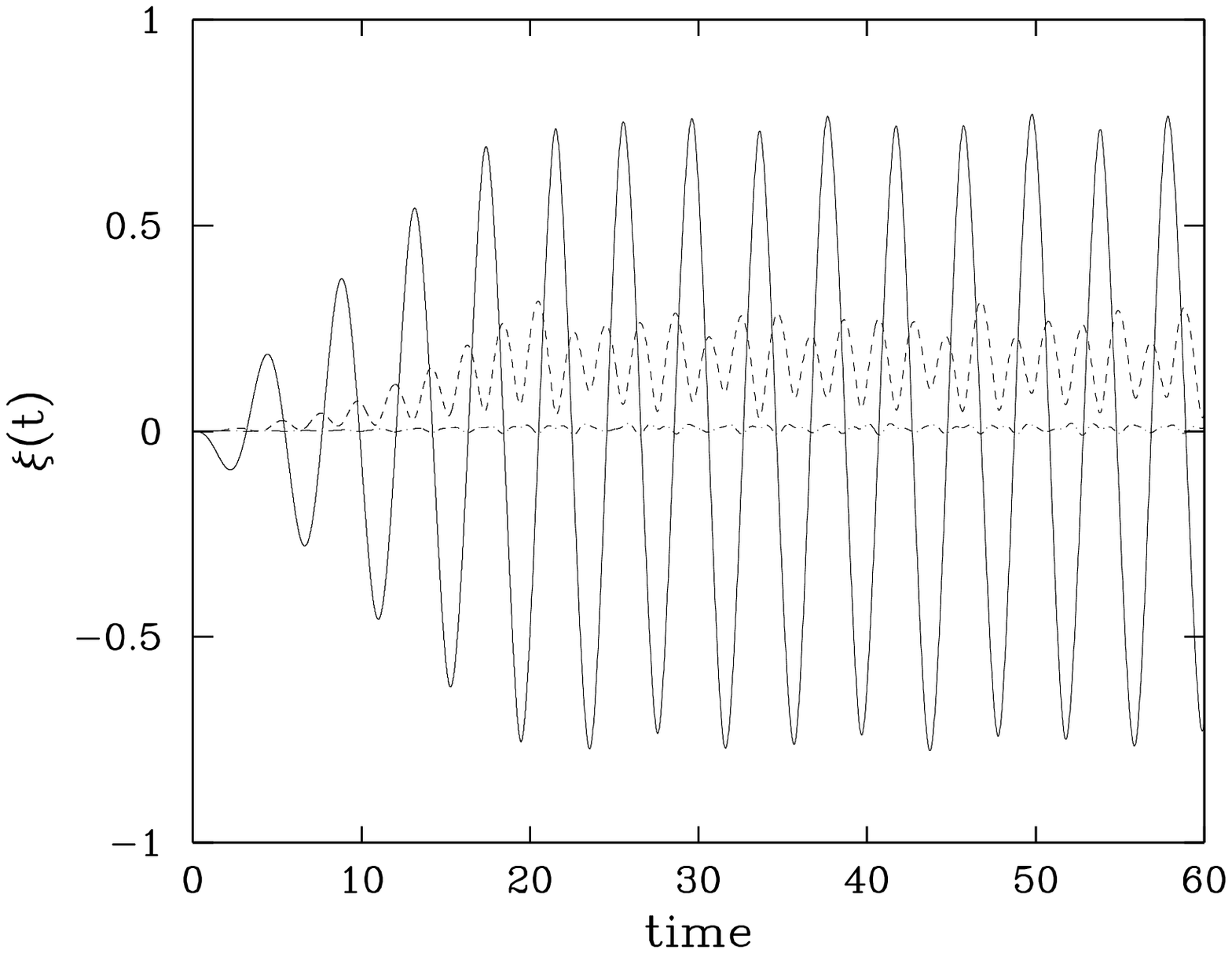,width=\linewidth}
  \begin{caption}
{  
Evolution in time of the three functions $\xi_2$ (solid line),
$\xi_-$ (dashed line) and $\xi_+$ (dot-dashed line) in a trap  with
$\lambda=\protect\sqrt{8}$. Time is in units 
of $\omega_\perp^{-1}$. In the limit
of small amplitude, these functions coincide with the $m=2$, low-lying
$m=0$ and high-lying $m=0$ normal modes, respectively. The driving 
force is the same as for the largest oscillation in 
Fig.~1.    } 
\end{caption}
\end{center}
\label{fig:m2-jila-xi}
\end{figure}

A strong enhancement of nonlinear effects can be obtained 
not only by increasing the strength of the driving force, but 
also by changing the anisotropy  of the trap. In fact, for 
special values of $\lambda = \omega_z / \omega_\perp$, frequencies 
of different modes, or of their harmonics, can coincide. 
An example is shown in Fig.~3, where we plot  
the functions $\xi_j(t)$ for an oscillation driven as in 
the case of the smallest oscillation in Fig.~1. 
The driving force is again tuned on $\xi_2$ with a relatively
small strength, but now the oscillation exhibits a strong nonlinear 
behavior. The crucial parameter is the value of $\lambda$, which 
in Fig.~3 is chosen to be $\sqrt{16/7}$.  As we
will discuss in the next sections, this special value of $\lambda$ is
associated with a resonance of the two modes $\xi_2(t)$ and 
$\xi_+(t)$; indeed, the figure shows an evident beating of these 
two modes.  
\begin{figure}
\begin{center}
  \epsfig{file=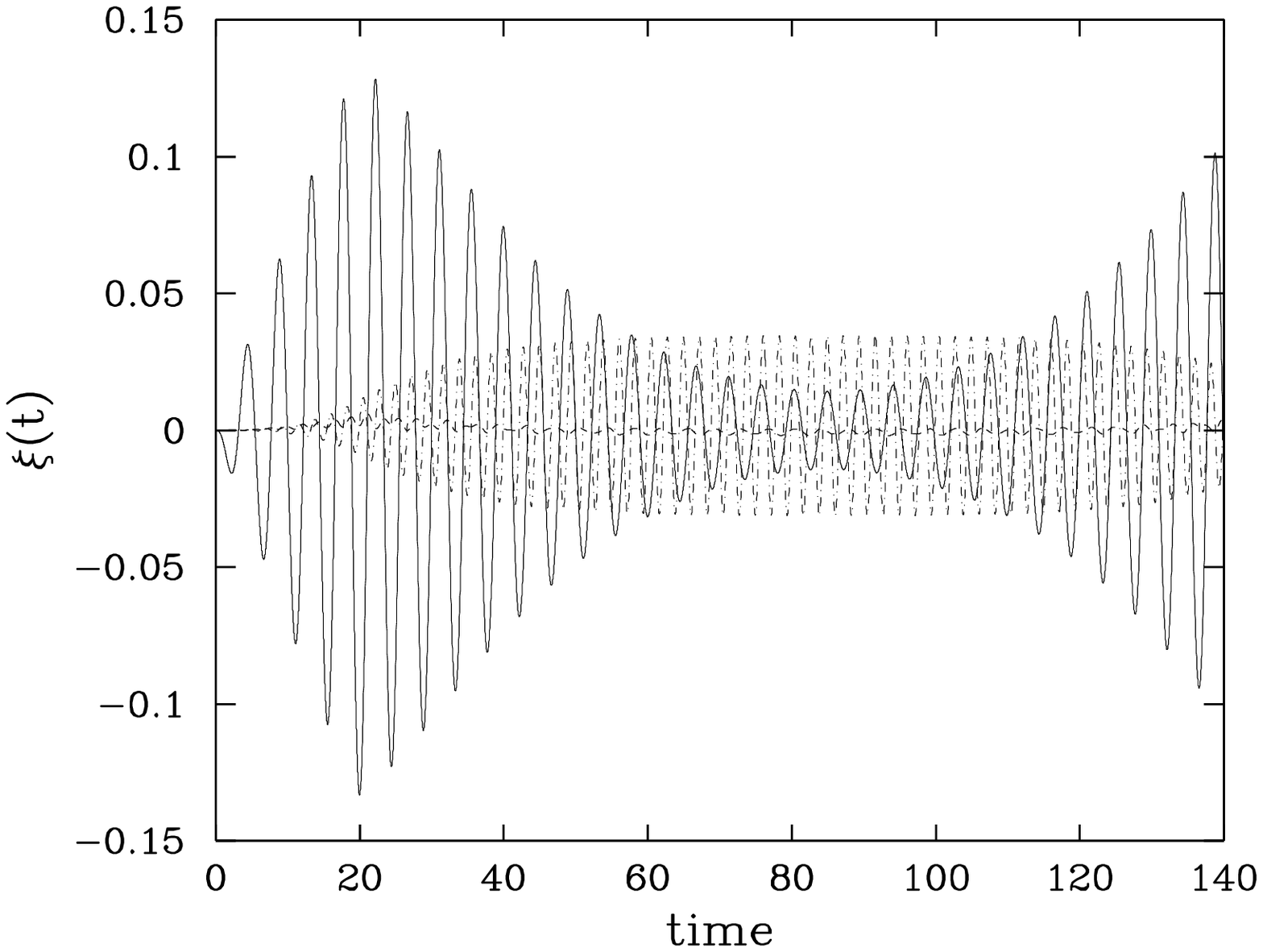,width=\linewidth}
  \begin{caption}
{  
Same as in Fig.~2 but for a trap having $\lambda=\protect\sqrt{16/7}$. 
The beating is between the $\xi_2$ mode, which is excited by the  
external force, and the $\xi_+$ mode, which is coupled through the 
second harmonic of $\xi_2$. The driving force is the same as for 
the smallest oscillation in Fig.~1. } 
\end{caption}
\end{center}
\label{fig:m2-res}
\end{figure}

At this point, before presenting further numerical results obtained 
from equations (\ref{ddotb}), we prefer to explore the first nonlinear 
corrections to the uncoupled normal modes. As we will see, an expansion 
of (\ref{ddotb}) at small amplitudes provides useful analytic formulae, 
which contain the main physical ideas and give accurate results.

\section{Analytic expansions at small amplitude}
\label{sec:analytic}

Equations (\ref{ddotb}), with $\omega_i=\omega_{0i}$ and 
$i=x,y,z$, describe free oscillations in the trap. They can 
be expanded by assuming $b_i(t)=1+\delta b_i(t)$ and keeping
the lowest orders in the small perturbation $\delta b_i$. The
linear terms provide the hydrodynamic spectrum of normal modes,
while harmonic generation is included in the higher order terms. One
can also use definitions (\ref{bx}-\ref{bz}) in order to get 
three coupled equations for the functions $\xi_-$, $\xi_+$ and
$\xi_2$, up to quadratic or cubic terms. We show here an example 
of an explicit calculation for the $m=2$ mode, while for the 
$m=0$ modes we will present only the final results. 

Let us assume the $m=2$ mode be excited in the cylindrically 
symmetric trap by some suitable driving  force, in such a way 
that $\xi_2(t) \approx A \cos(\sqrt{2}\ \omega_\perp t)$. If 
$A \ll 1$, the coupling with the other modes  is small and, hence, 
the $m=0$ oscillations $\xi_-$ and $\xi_+$ are expected to 
be of order $A^2$. Then, by expanding (\ref{ddotb}-\ref{bz}) up 
to third order in $A$, one gets the following equations:
\begin{equation}
\ddot{\xi}_2 + \omega_\perp^2 \xi_2 [ 2 - \xi_- q_- -\xi_+ q_+
+ 2 \xi_2^2] = 0 
\label{exp1}
\end{equation}
\begin{equation} 
\ddot{\xi}_\mp + \omega_\perp^2 q_\mp \xi_\mp \mp 
5 \omega_\perp^2 {\lambda^2 \over q_\pm} \left(
{ q_\pm -4 \over q_+ - q_- } \right) \xi_2^2 = 0 
\label{exp2}
\end{equation}
The two $m=0$ modes are not excited directly, but they 
can be driven by the $m=2$ mode {\it via} the last term in 
(\ref{exp2}). One  can easily find conditions for resonances: the
frequency of the $m=0$ mode, $\sqrt{q_\mp}\ \omega_\perp$, should be
equal to the frequency  of the second harmonic of the $m=2$ mode, 
$2\sqrt{2} \ \omega_\perp$. This never occurs  for the lowest 
$m=0$ mode, since $q_-$ is always smaller than $10/3$. Conversely,
for the high-lying mode $\xi_+$, a resonance is obtained for  
$\lambda=\sqrt{16/7}$. For this value of the asymmetry parameter, 
a small driving force, appropriate for exciting the $m=2$ mode, 
give rise to a coupled motion of both the $m=2$ and high-lying $m=0$ modes. 
This is exactly what we have already shown in Fig.~3
by solving the hydrodynamic equations (\ref{ddotb}) for the same 
special value of $\lambda$.  This enhancement of nonlinearity, due to 
resonances {\it via} second harmonics, has been recently 
suggested  also in  Ref.~\cite{Graham}.  Even though equations 
(\ref{exp1}-\ref{exp2})  are valid in the limit 
of small amplitudes, they give the correct conditions for  
the occurrence of strong nonlinear effects in the {\it exact} 
hydrodynamic solutions. 

The above equations can be also used to predict the frequency shift of 
the collective modes at the lowest order in the amplitude $A$.  In fact,
equations (\ref{exp2}) can be solved explicitly, by inserting $\xi_2$
at the leading order: $\xi_2(t) = A_2 \cos(\sqrt{2}\ \omega_\perp t)$. 
One finds
\begin{equation}
\xi_\mp (t) = { \pm (q_\pm -4) \over 2 (q_+-q_-) (q_\mp -8)} 
[q_\mp \xi_2^2(t) - 4 A_2^2]   \; . 
\label{alfabeta2}
\end{equation}
Using these solutions into (\ref{exp1}), one gets an equation for 
$\xi_2$ at the next order: 
\begin{eqnarray}
\ddot{\xi}_2 & + & 2 \omega_\perp^2 \left( 1 - A_2^2 {16 - 5 \lambda^2 
\over 16 - 7 \lambda^2 } \right) \xi_2 \nonumber \\
& + &  2 \omega_\perp^2 \left( 1
+ { 16 - 3 \lambda^2 \over 16 - 7 \lambda^2 } \right) \xi_2^3 = 0 \; . 
\label{gamma}
\end{eqnarray} 
This is a special case of the more general class of equations for 
anharmonic oscillators:  
\begin{equation}
\ddot{\xi} + \Omega_0^2 f + \gamma_2 \xi^2 +\gamma_3 \xi^3 = 0 \; . 
\label{anharmonic}
\end{equation}
One can easily prove that the solution $\xi(t)$ has a frequency 
$\omega$ which depends on the amplitude $A$ through
\begin{equation}
\omega = \Omega_0 \left( 1 - {5 \gamma_2^2 A^2 \over 12 \Omega_0^4 }
+ {3 \gamma_3 A^2 \over 8 \Omega_0^2 } \right) \; . 
\label{shift0}
\end{equation} 
By comparing (\ref{gamma}) and (\ref{anharmonic}) one identifies 
\begin{equation} 
\Omega_0^2 = 2 \omega_\perp^2 \left( 1 - A_2^2 {16 - 5 \lambda^2 
\over 16 - 7 \lambda^2 } \right)
\label{Omega}
\end{equation}
\begin{equation}
\gamma_3 = 2 \omega_\perp^2 \left( 1
+ { 16 - 3 \lambda^2 \over 16 - 7 \lambda^2 } \right)
\label{gamma3}
\end{equation}
and $\gamma_2=0$. Inserting these definitions into (\ref{shift0}), 
one gets finally
\begin{equation}
\omega = \sqrt{2} \ \omega_\perp \left[ 1 + { (16 - 5 \lambda^2) 
\over 4 (16 - 7 \lambda^2) } A_2^2 \right] \; ,
\label{shift-m2} 
\end{equation} 
which is the analytic expression of the frequency shift as a function 
of the anisotropy of the trap. Here the pathology at $\lambda=
\sqrt{16/7}$ is evident: as already said, it corresponds to a 
resonance induced by harmonic generation. 

In a more general form, one can write
\begin{equation}
\omega = \omega_0 [ 1 + \delta(\lambda) A^2]  \; , 
\label{shift}
\end{equation}
where $\omega_0$ is the  frequency  of each normal mode  
in the linear regime ($\sqrt{2} \omega_\perp$ and $\sqrt{q_\mp}
\omega_\perp$ for the $m=2$ and $m=0$ modes, respectively) while
$A$ is its amplitude.   The coefficient for the $m=2$ mode is then 
\begin{equation}
\delta_2 (\lambda) = { (16 - 5 \lambda^2) 
\over 4 (16 - 7 \lambda^2) } \; . 
\label{delta-2}
\end{equation}
Expansion (\ref{shift}) is expected to be reliable whenever
$| \delta | A^2 \ll 1$ and this excludes values of $\lambda$
too close to resonances. The coefficients $\delta_{\mp}(\lambda)$
for the $m=0$ modes can be calculated straightforwardly, with 
the same procedure used above for $m=2$. In particular, one
has to start exciting one of the two modes, $\xi_\mp$, in the
form $A_\mp \cos(\sqrt{q_\mp} \ \omega_\perp t)$; the other one is 
excited to the order $A_\mp^2$, while $\xi_2$ is never excited, due to
the axial symmetry of both the equations and the initial conditions. 
Expanding (\ref{ddotb}) up to
terms in $A_\mp^3$, one gets two coupled equations for $\xi_\mp$ of
the type (\ref{anharmonic}), while $\xi_2$ is decoupled. One 
finally obtains
\begin{eqnarray}
\delta_{\mp} & = & \frac{5}{2} \lambda^{2}
\frac{(q_{\pm}-2)(q_{\mp}-4)(q_{\mp}-5)}{(4 q_{\mp}-q_{\pm})
(q_{\pm}-q_{\mp})^{2}} \left[-1+\frac{15}{4}
\frac{\lambda^{2}}{q_{\mp}^{2}} \right]  \nonumber\\
& - & \frac{15}{16}\frac{1}{(q_{\pm}-q_{\mp})^{2}} \left[ -q_{\mp}+2 
\lambda^{2} q_{\mp}-9 \lambda^{2}+8 \right]^{2}  \nonumber\\
& - &  \frac{9}{4}\frac{(q_{\pm}-4)}{q_{\mp}(q_{\mp}-q_{\pm})}  
\nonumber\\
& - & \frac{3}{20}\frac{q_{\mp}-3}{q_{\mp}-q_{\pm}} \left[
-10 \lambda^{2} q_{\mp}+37 \lambda^{2}+11 q_{\mp}-54 \right] \; . 
\label{delta-m0} 
\end{eqnarray}
Two resonances are found by exciting the low-lying $m=0$ mode.
They occur when the frequency of the high-lying mode is equal
to the second harmonic of the low-lying mode; this happens for
$\lambda= (\sqrt{125} \pm \sqrt{29}) /\sqrt{72}$ (i.e., 
$\lambda \approx 0.683$ and $\lambda \approx 1.952$). 
Conversely, no resonances are found by exciting the high-lying 
$m=0$ mode. As an example of numerical values of $\delta_\mp$,  
let us consider a spherical trap, $\lambda=1$; in this case one
has $q_-=2$, $q_+=5$, $\delta_-= 11/12$ and $\delta_+ = -1/6$.
The latter value was already used in Ref.~\cite{Pitaevskii} to 
give an estimate of the collapse time for the high lying $m=0$ 
mode. 

The main results of these analytic expansions, together with 
the numerical solutions of (\ref{ddotb}), will be discussed 
in the following section.  
\begin{figure}
\begin{center}
  \epsfig{file=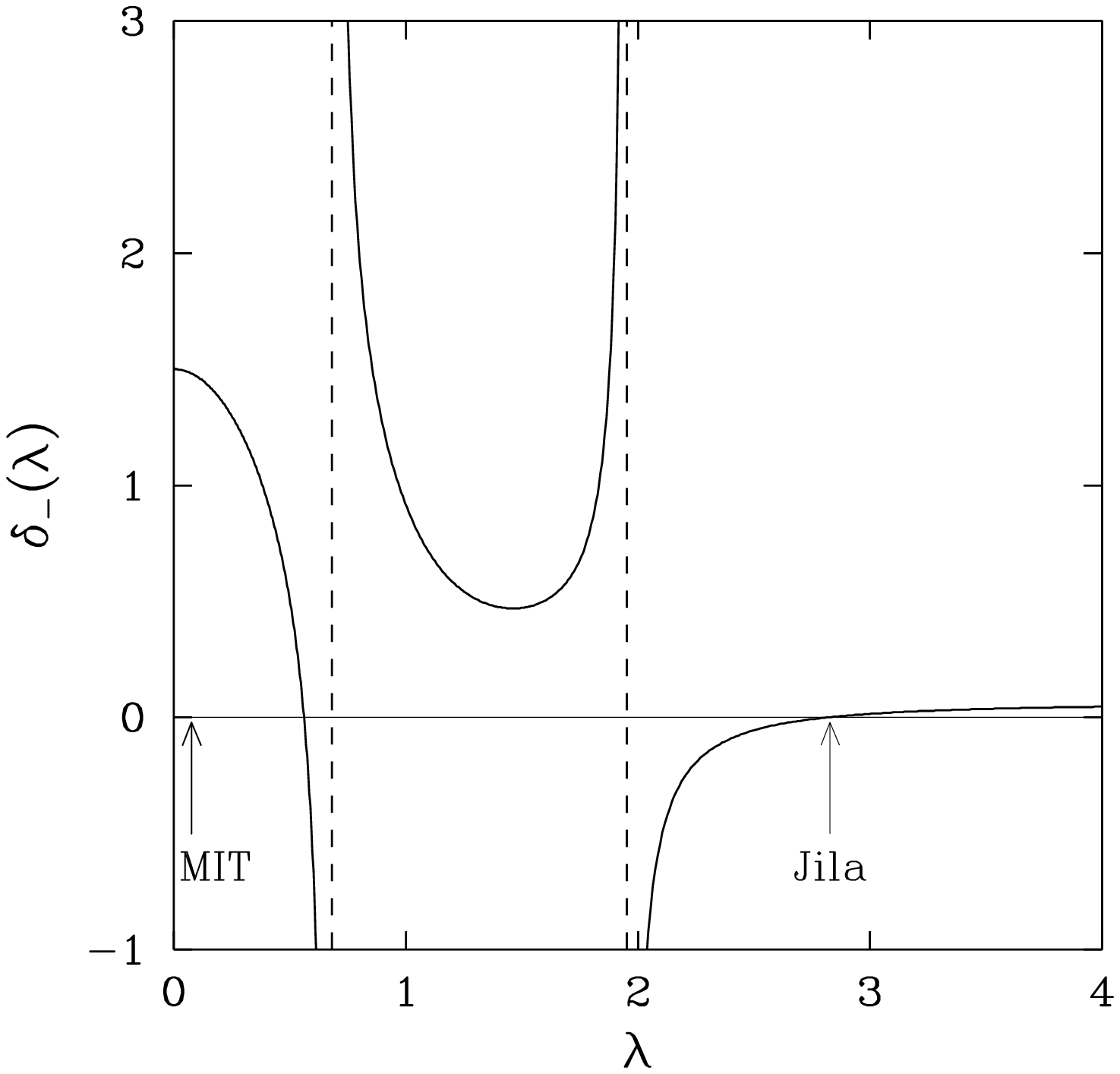,width=\linewidth}
  \begin{caption}
{  
 Coefficient of the quadratic expansion (\protect\ref{shift})
for the frequency shift of the low-lying $m=0$ mode, as a function 
of  the anisotropy parameter $\lambda=\omega_{0z}/\omega_\perp$.  
Divergences are found for $\lambda= (\protect\sqrt{125} \pm 
\protect\sqrt{29}) / \protect\sqrt{72}$. The values of $\lambda$ 
for the experimental traps of Refs.~\protect\cite{Jila,MIT} are
also indicated. } 
\end{caption}
\end{center}
\label{fig:resonances1}
\end{figure}
\begin{figure}
\begin{center}
  \epsfig{file=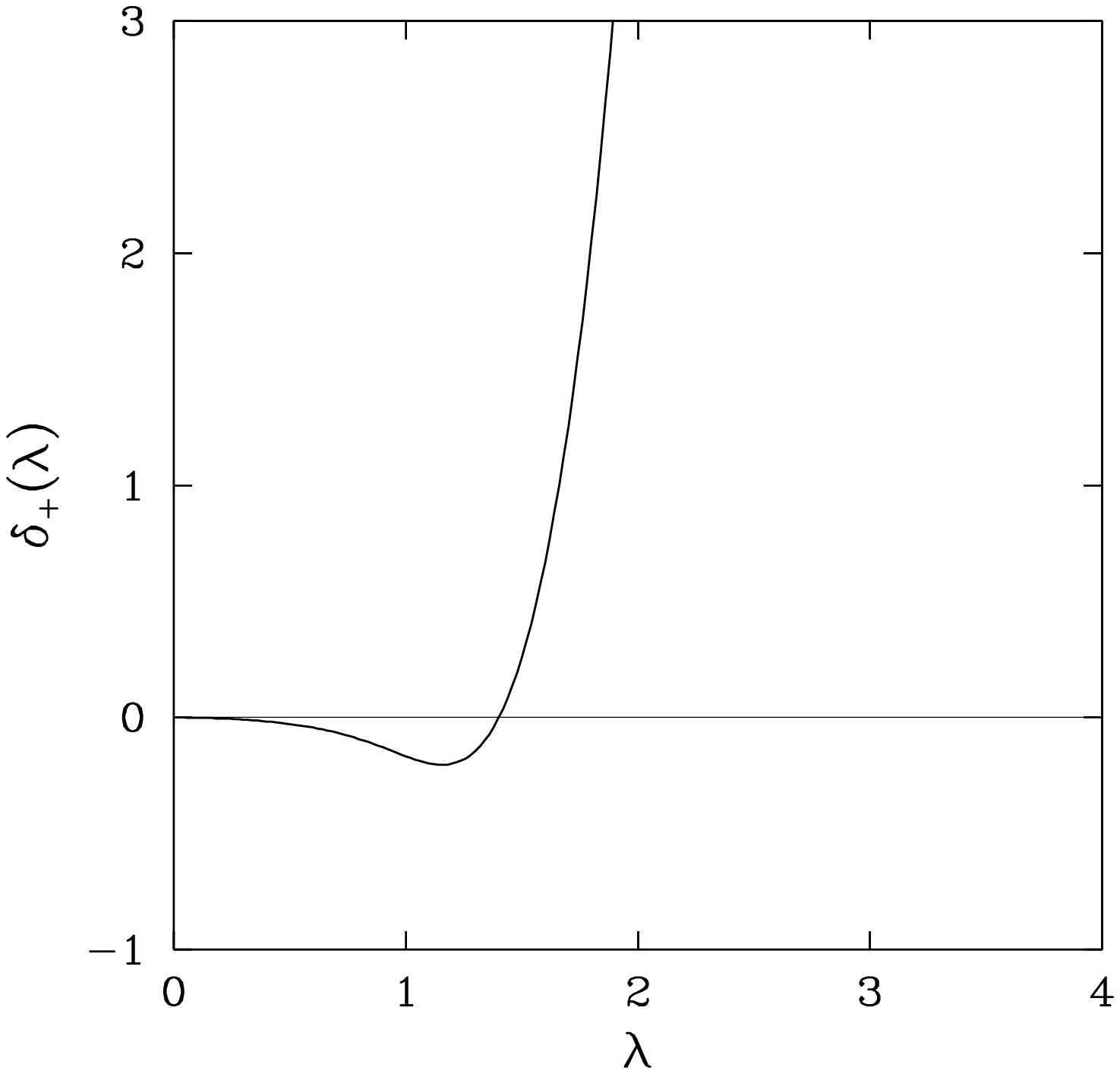,width=\linewidth}
  \begin{caption}
{  
Coefficient of the quadratic expansion (\protect\ref{shift})
for the frequency shift of the high-lying $m=0$ mode. For 
$ \lambda= \protect\sqrt{8}$ the coefficient is about $31$.  } 
\end{caption}
\end{center}
\label{fig:resonances2}
\end{figure}

\section{Frequency shift and mode coupling for different traps}
\label{sec:results}

So far have developed two methods for investigating the 
behavior of the $m=0$ and $m=2$ modes in a cylindrically 
symmetric trap. One is the numerical solution of the coupled
differential equations (\ref{ddotb}) for the widths $b_j(t)$,
which corresponds to solve exactly the equations of 
hydrodynamics for these collective modes. The other one is 
an expansion at small amplitude, which provides analytic 
solutions as well as simple formulae for the frequency shift.

In Figs.~4-6 we first 
show the three curves for the coefficients $\delta_- (\lambda)$, 
$\delta_+(\lambda)$  and $\delta_2(\lambda)$, from 
(\ref{delta-2}-\ref{delta-m0}), entering the quadratic expansion 
(\ref{shift}) of the frequency shift.  These
curves can be considered the main results of the present work.
The occurrence of resonances is a peculiar feature of the dynamics 
of the condensate. Close to these special values of $\lambda$ where
$\delta(\lambda)$ diverges, the frequency shift is expected 
to show an  anomalous enhancement and the time evolution of the 
widths of the condensate should become very irregular, due to 
significant mode coupling. 
\begin{figure}
\begin{center}
  \epsfig{file=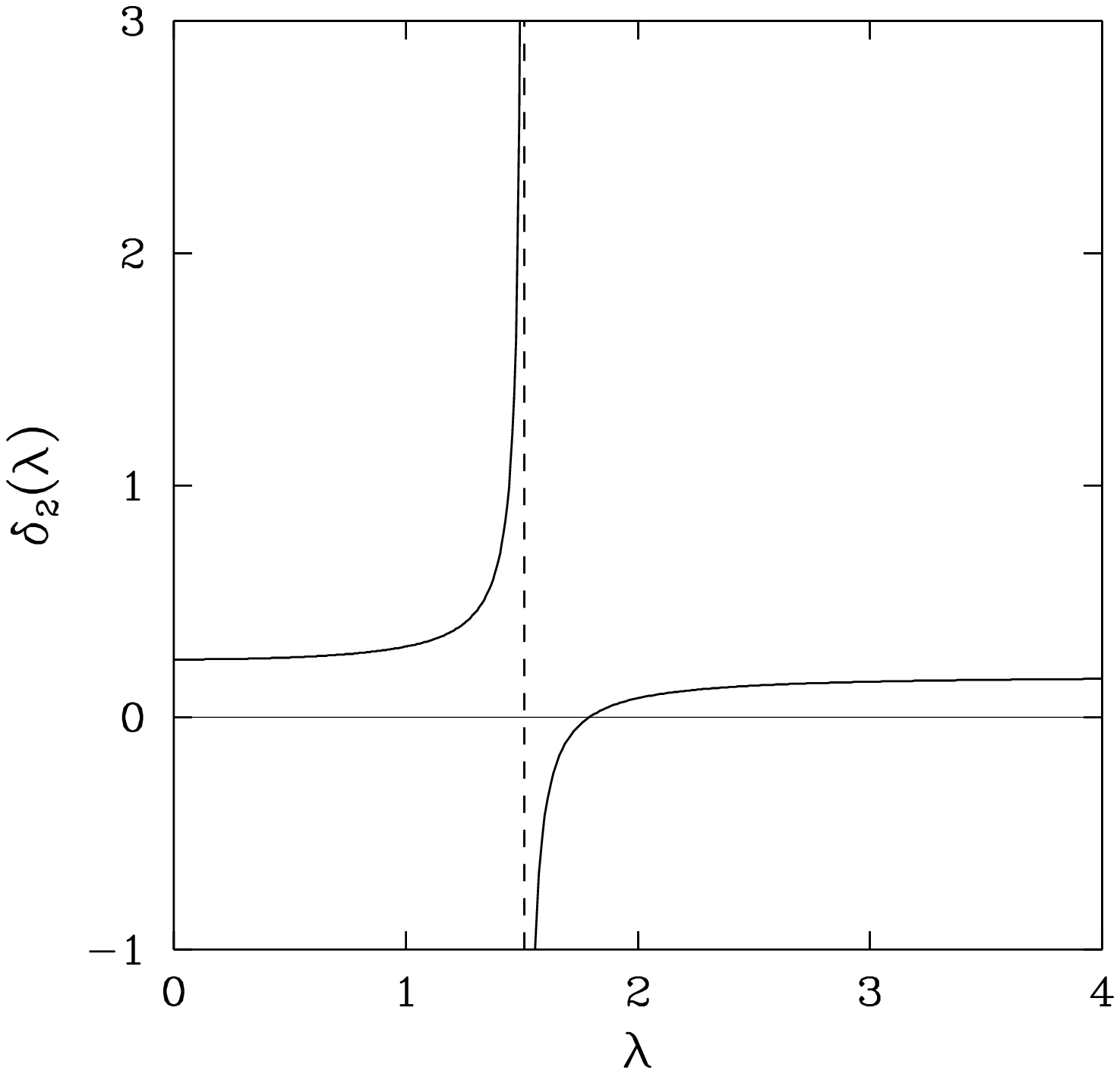,width=\linewidth}
  \begin{caption}
{  
Coefficients of the quadratic expansion (\protect\ref{shift})
for the frequency shift of the  $m=2$ mode. A divergence is 
found at $\lambda=\protect\sqrt{16/7}$. } 
\end{caption}
\end{center}
\label{fig:resonances3}
\end{figure}

In order to test the accuracy of the small amplitude expansion,
yielding the curves in Figs.~4-6, 
we compare the results obtained from both the numerical solution 
of (\ref{ddotb}) and the quadratic law (\ref{shift}) for specific
values of $\lambda$. In Fig.~7 we show the results 
for the $m=2$ and low-lying $m=0$ modes at $\lambda=\sqrt{8}$,
as in Jila experiments \cite{Jila}. In this case  one finds the 
coefficients $\delta_2 (\sqrt{8}) =3/20$ and $\delta_-(\sqrt{8})= 
1.636 \times 10^{-3}$. Solid lines are obtained from the numerical
solutions of (\ref{ddotb}), using definitions (\ref{bx}-\ref{bz}), 
and performing sinusoidal fits and/or Fourier analysis of $\xi_2$ and 
$\xi_-$. The frequency shift of the $m=0$ mode is  smaller than the 
one of $m=2$ especially for small amplitudes; in the $A \to 0$ limit,
the agreement between the numerical  results and the quadratic 
law (dashed lines) is very good.
In the $m=2$ case it remains good even for large amplitudes, 
while higher order corrections seem to be important for the $m=0$
mode at amplitudes larger than $0.2$  (corresponding to relative 
amplitudes of the order of  $20\%$ in the radial width of the 
condensate). In the latter case, one notices also that the 
coefficient $\delta_-$ is very small, since the value $\lambda=
\sqrt{8}$ is incidentally very close to a root of the function 
$\delta_-(\lambda)$, as shown in Fig.~4. 
Thus the first nonzero contributions to the shift come from 
powers $A^3$ or higher. We note also that the abscissa in 
Fig.~7  is the amplitude of $\xi_2(t)$ and 
$\xi_-(t)$  and not of the  width $b_x(t)$, which is the observed 
quantity at Jila. The reason is that the quadratic expansion  
(\ref{shift}) have analytic coefficients only for the normal 
modes $\xi_j$.  However, since $\lambda=\sqrt{8}$ is 
relatively far from resonances, the coupling between different 
modes in  (\ref{bx}-\ref{by}) is rather weak and the relative 
amplitude of  the oscillations of $b_x$ practically coincides 
with the amplitude of each  normal mode, within the range of 
Fig.~7.
\begin{figure}
\begin{center}
  \epsfig{file=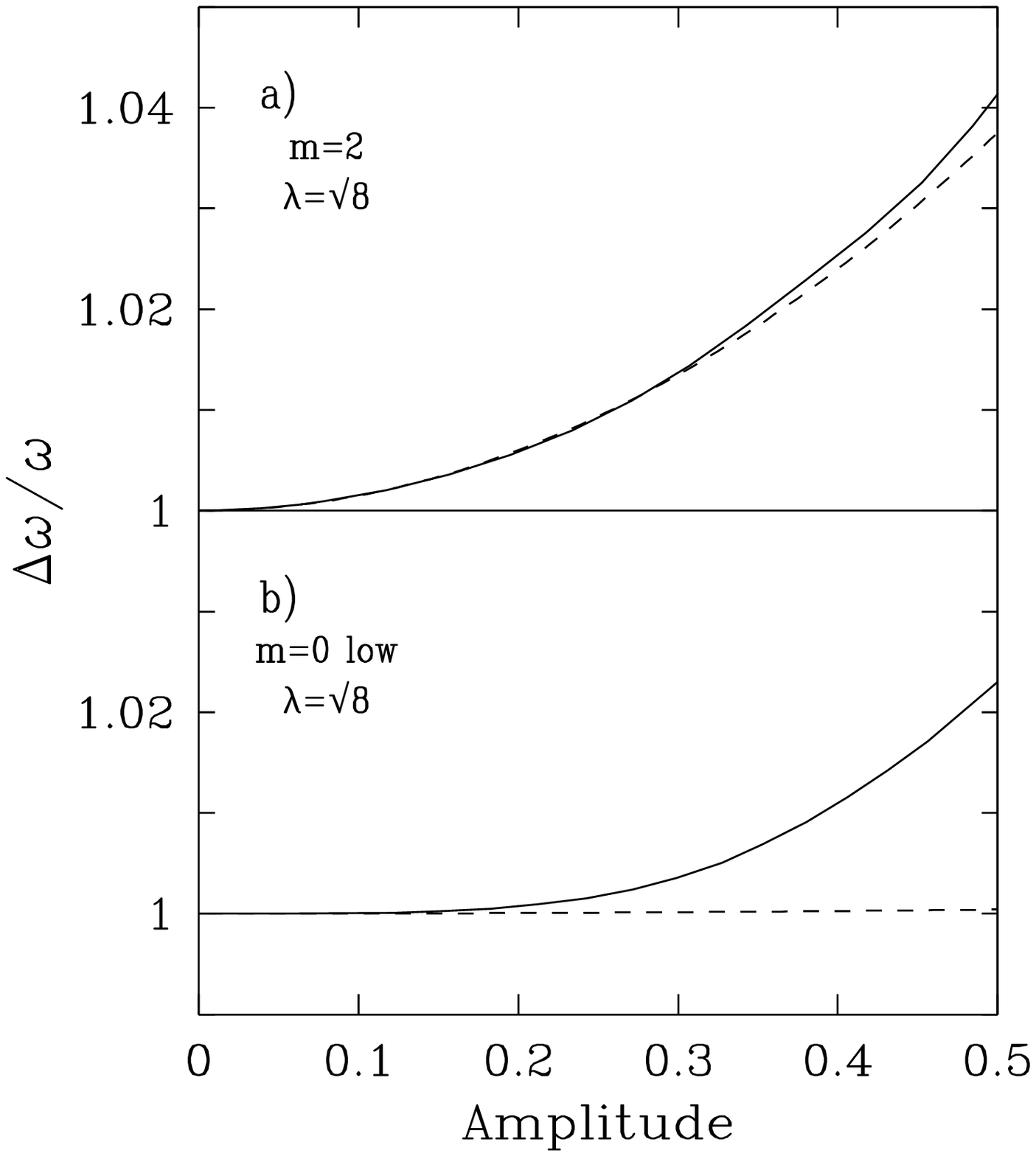,width=\linewidth}
  \begin{caption}
{  
Frequency shift for the $m=2$ and low-lying $m=0$ modes
in the Jila trap as a function of their amplitude. 
Solid line: from the numerical solution of
(\protect\ref{ddotb}); dashed line: quadratic expansion 
(\protect\ref{shift}) with $\delta_2 = 3/20$ and $\delta_-
= 1.636 \times 10^{-3}$ from (\protect\ref{delta-2}-
\protect\ref{delta-m0}). } 
\end{caption}
\end{center}
\label{fig:shift-jila}
\end{figure}

It is not easy to compare these results with the available data 
for the frequency shift in the Jila experiments \cite{Jila}. The
data show no significant frequency shift for the low-lying $m=0$ 
mode and a relatively large shift for the $m=2$ mode. The fact that 
the shift is larger for $m=2$ than for $m=0$ is in agreement with 
our predictions, but a quantitative comparison 
should take into account at least two further effects. First,  the 
data have been taken after switching off the trapping potential. As 
discussed in our previous work \cite{TN} (see also \cite{note}), 
the expansion  of the gas causes an amplification of the relative 
amplitude of the observed oscillations.  
This amplification is different in the three 
directions, depending  on the anisotropy of the trap and on the 
excited modes.  Second, the assumption  $Na/a_{HO} \gg 1$, which is at 
the basis of the hydrodynamic approach,  is not well satisfied 
by the samples used so far at Jila. In this case,  the role of 
the quantum pressure, beyond the Thomas-Fermi approximation  
could be relevant also for the dynamics of the expansion.   

In Fig.~8 we plot the frequency shift of the 
low-lying $m=0$ mode for the MIT trap \cite{MIT} with 
$\lambda=0.077$. Again the solid line comes from the evolution 
of $\xi_-(t)$, calculated numerically from (\ref{ddotb}), 
while the dashed line is the quadratic approximation 
(\ref{shift}), whose coefficient is now $\delta_- = 1.481$. 
The agreement between the two curves is good.
In this trap, the condensate is cigar-shaped and one
measures the oscillations of the axial width $b_z(t)$. Looking
at equation (\ref{bz}), one notes that the relative amplitude 
of the oscillations of $b_z$ is about four times the amplitude
of the normal mode $\xi_-$. In fact, for such a small value 
of $\lambda$, the quantity $q_-$ behaves like $(5/2)\lambda^2$
and, hence, can be neglected in (\ref{bz}); moreover, the 
coupling with the high-lying $m=0$ mode is weak and thus one
can also neglect $\xi_+$ at the lowest order. This means that
the shift $\delta_- A^2$ becomes $(\delta_- /16) A_z^2$, if
$A_z$ is the relative amplitude of the axial width $b_z$. The shift
in frequency is thus very small; for instance, when the axial 
width  oscillates with a relative amplitude of $40\%$, the 
amplitude of $\xi_-$ is about $0.1$ and the predicted 
shift is less than $2\%$. Actually the oscillations in Ref.~\cite{MIT}, 
which do not exhibit any frequency shift, have been imaged after 
the free expansion of the condensate and correspond to relative 
amplitudes of the  order of $10\%$, or less, for $b_z$  before 
the expansion. Therefore,  the predicted 
shift is practically zero within the accuracy of the available 
experiments and the $A^2$ law can not be tested. The situation
can be greatly improved by the use of nondestructive imaging 
techniques,  as in the trap of Ref.~\cite{MIT2}.

As concerns nonlinear coupling of $m=0$ and $m=2$ modes, the 
values of $\lambda$ in the Jila and MIT traps are not of 
particular interest. The curves 
in Figs.~4-6
suggest different choices, namely $\lambda=\sqrt{16/7}$ for
the $m=2$ mode and $\lambda= (\sqrt{125} \pm \sqrt{29}) /\sqrt{72}$
for the low-lying $m=0$ mode.  It is also worth mentioning
that the curves in  Figs.~4-6
come from an expansion  
up to the third order in $A$, as in (\ref{exp1}-\ref{exp2}),
and the coupling occurs {\it via} the second harmonics. When the 
amplitude is large, one expects higher order terms  to become 
significant, providing other special values of $\lambda$ for
mode coupling. For instance, the hydrodynamic 
equations (\ref{ddotb}) predict a beating of the $m=2$ and the 
high-lying  $m=0$ modes for $\lambda=\sqrt{63/11}$. This beating 
can be calculated analytically by extending equations 
(\ref{exp1}-\ref{exp2}) to the next order in $A$; for that value of
$\lambda$,  one then finds that the third harmonic of $\xi_2$ 
has the same frequency  of $\xi_+$.  
Finally, we remark  that other values of $\lambda$ can 
give rise  to similar effects, even without harmonic generation. This
happens when an accidental degeneracy occurs in the spectrum 
of normal modes, including $|m|>2$, as recently discussed by 
\"Ohberg {\it et al.} \cite{Ohberg}. 
\begin{figure}
\begin{center}
  \epsfig{file=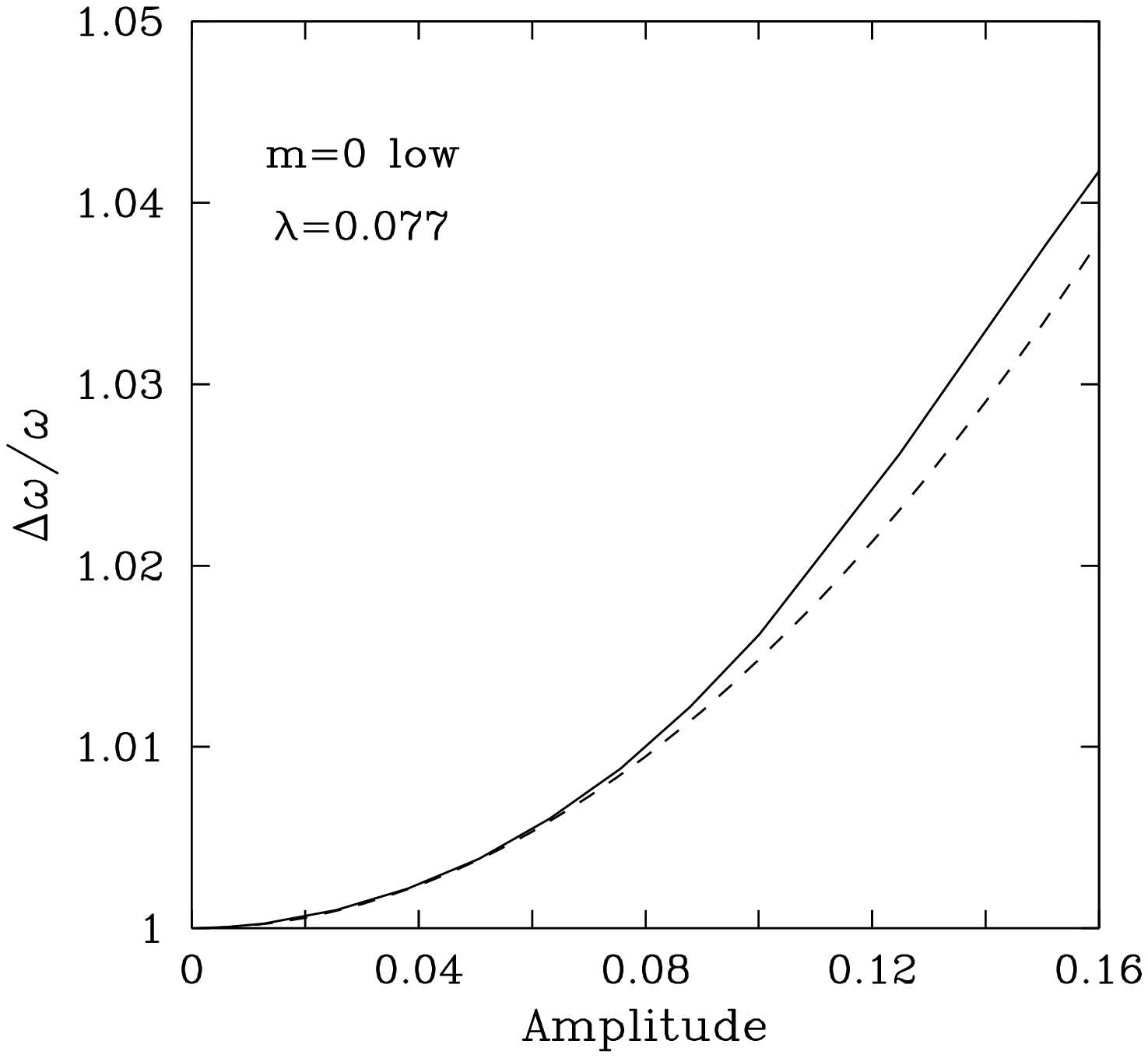,width=\linewidth}
  \begin{caption}
{  
Frequency shift for the  low-lying $m=0$ mode
in the MIT trap as a function of its amplitude. 
Solid line: from the numerical solution of
(\protect\ref{ddotb}); dashed line: quadratic expansion 
(\protect\ref{shift}) with $\delta_- = 1.481$ 
from (\protect\ref{delta-m0}).} 
\end{caption}
\end{center}
\label{fig:shift-mit}
\end{figure}

\section{Application to the theory of collapse and revival}
\label{sec:collapse}

The fact that the frequency of the normal modes depends on
their amplitude has  interesting consequences for the phenomenon 
of collapse  and revival of collective excitations.  The latter 
belongs to a class of quantum phenomena having no classical analogue. 
Examples are the collapse and revival of coherent quantum states 
observed in atomic Rydberg wave packets \cite{at}, in molecular 
vibrations \cite{mol} and for atoms interacting with an
electromagnetic field in a resonant cavity \cite{2} (see 
\cite{ave,the} and references therein, for a theoretical 
discussion).  The same concepts have been recently applied to 
the trapped condensate (see refs.~\cite{Pitaevskii,bir} and
references therein). 
In this confined Bose system, the collapse of collective 
excitations originates from a dephasing during the oscillation.
In Ref.~\cite{bir} the dephasing was associated with the 
fluctuations  of the particle occupation number and the collapse 
time was estimated within Hartree approximation. One of 
us \cite{Pitaevskii} has 
recently developed a theory for the same process, but based on 
the fluctuations in the number of quanta of oscillation. The main 
idea is that a collective mode is, in general, a coherent superposition
of stationary states of an oscillator and the number of quanta of 
oscillations is the natural quantity for classifying those states. 
The estimated time scale for the collapse turns out to depend, as a key 
ingredient, on the frequency shift of the normal modes. We will recall 
here the main steps of the theory and then we will use our new results 
for the frequency shifts in order to provide a quantitative estimate 
of the collapse time. 

As we saw in section \ref{sec:analytic}, the collective variable $\xi$ 
can be considered as solutions of the equations for anharmonic 
oscillators. Let us write the frequency, up to the first non linear 
correction, in the form 
\begin{equation}
\omega = \omega _0+\delta \omega = \omega _0 (1+\kappa E) ,
\label{1}
\end{equation}
where $E$ is the energy of the oscillations. We assume here the 
condition of weak nonlinearity, $| \kappa | E \ll 1$, even though 
the theory can be generalized to the case of large amplitude. In order to 
provide a quantum description, one can use the semiclassical expression 
$\hbar \omega = (\partial  E_n/\partial n)$, which allows one to to 
rewrite equation (\ref{1})  in the quantum form  $\omega _n = 
E_n/\hbar=\omega _0n +bn^2/2$. Here the quantity $n$ is the number of 
quanta ($n \gg 1$) in a given excited state of energy $E_n$, while
the coefficient $b$ is given by $b=\hbar \omega _0^2 \kappa$.
The oscillations in the experiments \cite{Jila,MIT} were driven
through a sinusoidal force into a coherent state of the oscillator. 
The wave function of a such a stationary state can be written in the 
form
\begin{equation}
\psi = \sum _n c _n \psi _n \exp (-i\omega _nt) \; ,
\label{psi}
\end{equation}
where
\begin{equation}
\mid c _n \mid ^2 =   { \bar{n}^n \over n! } \exp(-\bar{n})
\approx \frac{1}{\sqrt{2\pi
\bar{n}}} \ \exp \left[ -\frac{(n-\bar{n})^2}{2\bar{n}} \right] \; ,
 \label{4}
 \end{equation}
where $\bar{n}$ is the average number of quanta ($\bar{n} \gg 1$) and
$E=\hbar \omega _0 \bar{n}$.  Let us now consider the oscillator 
co-ordinate $\xi(t)$ and calculate its average over the
state (\ref{psi}-\ref{4}).  If one considers only $n \rightarrow 
n \pm 1$ transitions , the result is $\langle \xi (t) \rangle 
\propto  \sum _n\mid c _n \mid ^2 \cos [(\omega _0 +bn)t]$.  
For small enough values of $t$,  one can replace the summation 
over $n$ by an integral and one gets a Gaussian damping of the 
oscillation according to $\langle \xi \rangle \sim \exp(-\bar{n} b^2t^2/2) 
\equiv \exp[-(t/\tau _c)^2]$,  where
\begin{equation}
\tau_c ^{-1} \ = \ (\bar{n}/2)^{1/2} \mid b \mid \ = \
      \omega _0  (E \hbar \omega _0/2)^{1/2} \mid \kappa \mid  \; . 
 \label{8}
 \end{equation}
Since the amplitude of the oscillations is proportional to 
$\sqrt{ \bar{n}} $, the resulting amplitude dependence 
of the collapse time, $\tau _c $, is the same as in the theory of 
Ref.~\cite{bir}. The periodicity of $\langle \xi (t) \rangle$ gives 
also the revival period $\tau _r = 2\pi/(\hbar \omega _0^2 | \kappa | 
 )$ as in Ref.~\cite{ave}. The meaning of $\tau_c$ and $\tau_r$ 
is also shown schematically in Fig.~9. From the 
expressions for $\tau_c$ and $\tau_r$ one easily sees that 
$\tau_c \approx \sqrt{(1/\bar{n})} \ \tau_r \ll \tau_r$.  Note that 
in quantum mechanics the measurement of an oscillator coordinate $\xi$ 
is, generally speaking, destructive because it affects the oscillator 
momentum $p_{\xi}$. In other words, it  modifies the coefficients
$c_n$, thus  preventing the observation of collapse and revival.
In order to measure properly $\langle  \xi (t) \rangle$,  one 
must repeat cycles of observations with different replica of the 
system in different instants of time. This is, however, exactly 
the procedure used in the available  experiments \cite{Jila,MIT,cor1}.
\begin{figure}
\begin{center}
  \epsfig{file=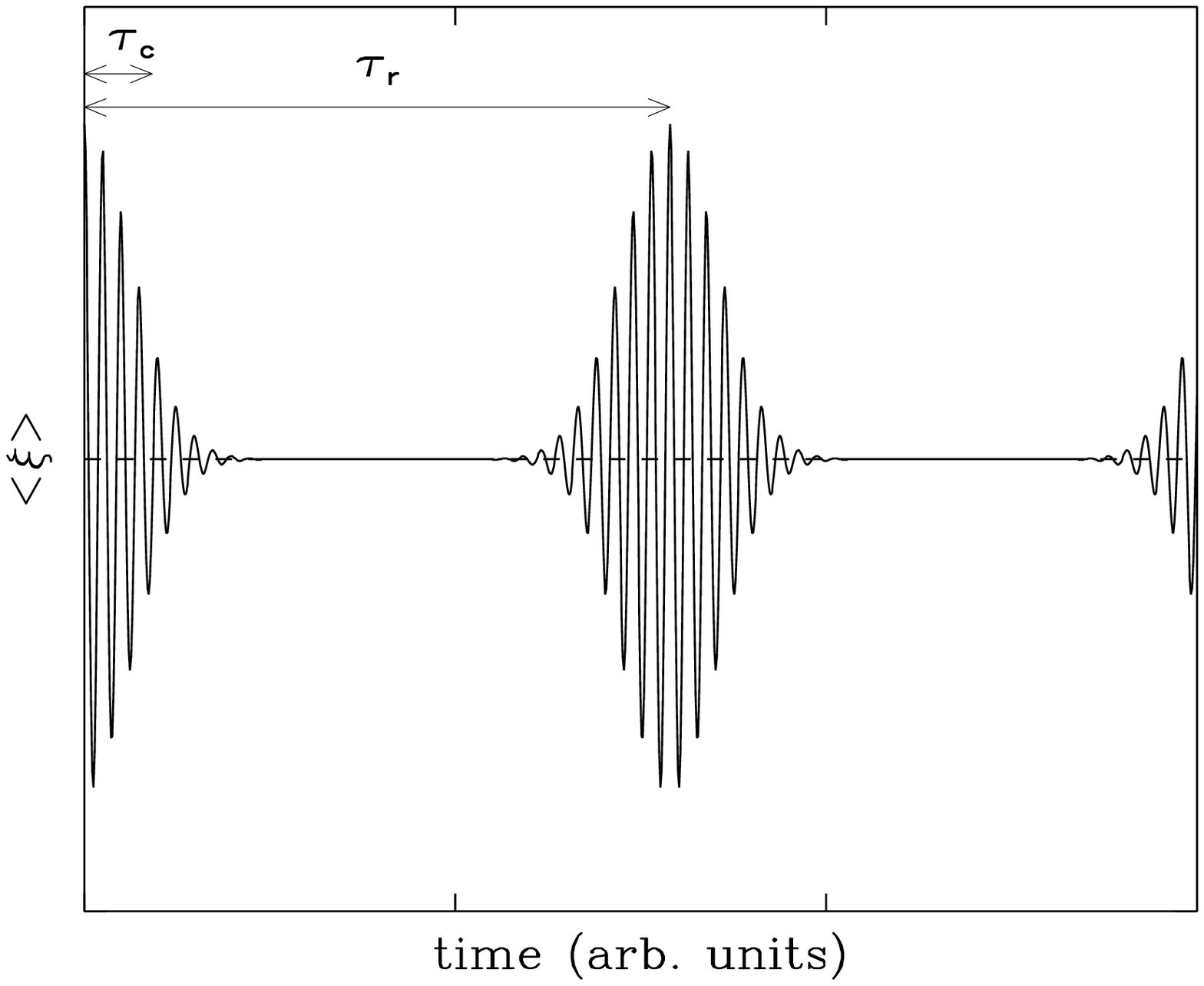,width=\linewidth}
  \begin{caption}
{ 
Schematic picture of collapse and revival. Both time and
$\langle \xi \rangle$ are in arbitrary units and the oscillations are 
just a pictorial view of the phenomenon. The two time-scales, $\tau_c$
and $\tau_r$ are indicated.  }  
\end{caption}
\end{center}
\label{fig:collapse}
\end{figure}

According to (\ref{8}) the collapse time $\tau _c$ decreases 
when  $\bar{n}$ increases. However, one has to keep in mind that 
the theory is restricted to the weak nonlinear regime 
$|\kappa | E \ll 1$, which corresponds to $\bar{n} \ll 
1/(\hbar \omega _0 | \kappa  |)$.  Thus, the collapse time 
is subject to the condition   $\tau _c \gg \tau _{min} 
\equiv  1/(\omega _0 \sqrt{\hbar \omega _0  |\kappa  | })$
and the collapse can be observed only if $\tau _{min} $ is shorter 
than the typical time scale for the dissipative damping of the 
oscillations, which is always present at finite temperature.

In order to provide quantitative estimates of the collapse time for
trapped Bose condensed gases,  one must calculate the 
energy of oscillations, $E$, as a function of their amplitude. 
This is what we have done in the previous sections. One has only to 
write the proper relation between  the nonlinearity  coefficients 
$\kappa $ and $\delta$.  To this aim, it is enough to calculate 
the energy $E$ in linear approximation, as twice  the mean kinetic
energy $E=m\int \! d{\bf r} \rho _0 v^2$, where $\rho _0({\bf r})$ is 
the equilibrium density of condensate and ${\bf v}({\bf r})$ is the 
velocity field associated with each normal mode. Direct integration 
for $m=2$ gives:
\begin{equation}
E=\frac{4}{7}\mu N A_2^2 \; ,
\label{E2}
\end{equation}
where $A_2$ is the amplitude of the mode and $\mu$ is the chemical 
potential of the system, calculated within the Thomas-Fermi 
approximation.  Similarly, for the $m=0$ modes one finds
\begin{equation}
E= \frac{1}{7} \left[ 2q_\mp  +(q_\mp -4)(3q_\mp -10) \right] 
\mu N A_\mp^2 \; ,
\label{E0}
\end{equation}
where the quantities $q_\mp(\lambda)$ are defined in (\ref{qmp}). 
The above energies can be rewritten in a more compact form as 
$E=\epsilon \mu A^2$,
where $\epsilon(\lambda)$ is the appropriate coefficient for each 
normal mode in (\ref{E2}-\ref{E0}). Inserting this result in 
(\ref{1}) and comparing with the quadratic law (\ref{shift}), one
finds
\begin{equation}
\kappa = { \delta  \over \epsilon \mu N} \; . 
\label{dim}
\end{equation}
The final result for the collapse time, from (\ref{8}), is then 
\begin{equation}
{\omega _0 } \tau _c  \ = \ { \sqrt{2} \ \epsilon \mu N \over  \mid 
\delta \mid  \sqrt{E\hbar \omega_0}}
\ = \  \frac{\sqrt{2\epsilon }}{A \mid \delta \mid }
\sqrt{\frac{\mu N}{\hbar \omega _0}}.
\label{gen}
\end{equation}

Let us estimate $\tau _c$ for the $m=2$ mode, assuming typical 
experimental parameters of  Ref.~\cite{Jila}: $N=4500$, 
$\omega _\perp /2\pi = 132$ Hz, $\lambda = \sqrt{8}$ and $a=110$ 
Bohr radii.  With these parameters one finds $\sqrt{\mu N/\hbar 
\omega_\perp} \approx 195$.  The frequency of the $m=2$ mode
in the linear limit is $\omega _0 = \sqrt{2} \omega _\perp$, while
the coefficient of the quadratic expansion (\ref{shift}) is
$\delta_2= 3/20$. Equation (\ref{E2}) gives $\epsilon _2 = 4/7$.
With a reasonable choice for the amplitude, $A=0.2$, the final 
result for the collapse time is $\tau_c=4.98$ s.
The authors of Refs.~\cite{Jila,cor1} reported a lifetime of 
the $m=2$ mode of the order of $100$ ms. It means that, under the
experimental conditions, the dissipative  damping is too strong 
and the collapse can not be observed.  It is not hopeless, however, 
to discover this effect under appropriate conditions. The point is 
that, according both to measurements \cite{cor1} and to  
theoretical considerations \cite{dam}, the dissipative damping 
decreases rapidly by lowering the temperature of the gas.
We believe that experiments at lower temperature  would permit 
to observe the quantum collapse of collective modes in these
macroscopic  objects. It is worth mentioning, in this
connection, the recent suggestion to cool adiabatically  the gas 
by changing the trapping frequency \cite{ad}.

The low-lying $m=0$ mode, which also has been observed,  exhibits a very 
small frequency shift and is not, for this reason,  a proper object to 
search for collapse, at least within the range of amplitude where the 
quadratic expansion (\ref{shift}) holds. Conversely, the high-lying $m=0$ 
mode looks  more  promising.  For this mode, using again the 
parameters of the Jila trap, one has $\omega _0 \approx 4.98\omega _\perp$, 
$\epsilon \approx 198$ and $\delta_+ \approx 30$. The oscillations 
are strongly anisotropic and  the amplitude in the $z$-direction is 
larger that in the radial one. According to (\ref{bz}) one gets
$A_z \approx (q_+-4)A_+ \approx 20.8A_+$.   By assuming again  
$A_z=0.2$, the collapse time becomes $\tau_c=1.4$ s.

From these results, it appears evident that the {\it resonances}
described in the previous sections of this work are very promising
even for the observation of collapse and revival of the oscillations. 
By a proper choice of the anisotropy parameter $\lambda$ one can
in fact increase significantly the coefficient $\delta$ and, 
consequently, lower the collapse time to an observable scale. 
Obviously, for real traps, one should also take into account possible 
nonlinear effects originating from trivial nonharmonic corrections 
to the magnetic confining potential. But the present scenario 
for collapse and revival seems plausible and, in any case, the first 
step for its confirmation should be the observation of the predicted 
amplitude dependence of the frequency shift.

\section{Conclusions}

In this paper we have investigated the behavior of collective excitations
of trapped Bose gases in the limit of zero temperature. The formalism of
hydrodynamic equations is suitable for deriving both numerical and 
analytic results, valid in the limit $Na/a_{HO}\gg 1$. As already shown by 
other authors, those equations describe the lowest collective excitations
in the linear regime (normal modes), in agreement with available
experiments. They can be also used for the dynamics of the system 
in nonlinear regime. 
We have shown here that they provide nontrivial predictions for the frequency 
shift of the normal modes. We have discussed the case of $m=0$ and $m=2$
modes in a generic cylindrically symmetric trap and we have explored the
peculiar behavior  of the condensate for special choices of $\lambda =
\omega_{0z}/\omega_\perp$, the asymmetry parameter of the trap. For those
special traps, the frequency of an oscillation becomes equal to the 
second harmonic of another one. This degeneracy enhances nonlinear effects,
producing significant consequences in measurable quantities: for 
instance, the time evolution of condensate shape can show  
irregular patterns and the frequency shift can become rather large. 
The physical picture has been supported by simple analytic results.  
We have also discussed an application to the the theory of 
collapse and revival of the collective oscillations.


\begin{references}
\bibitem{note} We remark here that the shift calculated in  
Ref.~\protect\cite{TN}  was affected by the inadequate 
procedure used for extracting the frequency of the oscillations. 
In that work, we used the peak-to-peak time interval, which however 
can depend on the coupling between different modes 
and on their relative phases.  The Fourier analysis and the 
sinusoidal fits, used in the present  work, are much safer methods. 
One finds significant differences only for the shift of 
the $m=2$ frequency, plotted in Fig.~4 of  Ref.~\protect\cite{TN}. 
The actual curve is approximately a factor $2$ lower, increasing 
the discrepancy with the available experimental data at large 
amplitude.
\end{references}
\end{document}